\documentclass[prb,twocolumn,showpacs]{revtex4}

\usepackage{graphicx}
\usepackage{dcolumn}
\usepackage{amsmath}

\begin{document}
\title{Interplay of superexchange and orbital degeneracy in Cr-doped LaMnO$_{3}$}
\author{Joachim Deisenhofer, Michalis Paraskevopoulos, Hans-Albrecht Krug von Nidda,\\
 and Alois Loidl}
\affiliation{Experimentalphysik V, Elektronische Korrelationen und
Magnetismus, Institut f\"{u}r Physik, Universit\"{a}t Augsburg, D
- 86159 Augsburg, Germany}
\date{\today}

\begin{abstract}
We report on structural, magnetic and Electron Spin Resonance
(ESR) investigations in the manganite system
LaMn$_{1-x}$Cr$_{x}$O$_{3}$ ($x\leq 0.5$). Upon Cr-doping we
observe a reduction of the Jahn-Teller distortion yielding less
distorted orthorhombic structures. A transition from the
Jahn-Teller distorted O$'$ to the pseudocubic O phase occurs
between $0.3<x<0.4$. A clear connection between this transition
and the doping dependence of the magnetic and ESR properties has
been observed. The effective moments determined by ESR seem
reduced with respect to the spin-only value of both Mn$^{3+}$ and
Cr$^{3+}$ ions.
\end{abstract}



\maketitle

\section{Introduction}

It was recently shown that in La$_{1-x}$Sr$_{x}$MnO$_{3}$ the
realization of an insulating and ferromagnetic ground state for
$0.1<x<0.15$ results from a superexchange (SE) driven
rearrangement of the high-temperature orbital order (OO)
established by the Jahn-Teller (JT)
distortion.\cite{Endoh99,Paraskevopoulos00b} Upon Sr-doping the JT
distortion in the parent compound gets suppressed and the
degeneracy of the Mn$^{3+}$ e$_{g}$ electrons becomes almost
restored. In this case the two-orbital model by Kugel and Khomskii
becomes valid and ferromagnetic SE interactions between Mn$^{3+}$
ions show up.\cite{Kugel82} These drive the system into an
orbitally ordered and ferromagnetic state. The exact nature of the
OO is still unclear, although some proposals have been made by
Khomskii recently,\cite{Khomskii01} based on the idea of the so
called orbital polaron by Kilian and Khaliullin. \cite{Kilian99}
The realization of a ferromagnetic and insulating state has been
recently adressed by Khomskii and Sawatzky \cite{Khomskii97} and
also by Mizokawa \textit{et al.}~\cite{Mizokawa00b} For even
higher Sr-doping levels ($x\geq 0.17$) the increasing
double-exchange (DE) interactions become dominant and establish a
ferromagnetic and metallic ground state.

Since the first studies of the series LaMn$_{1-x}$Cr$_{x}$O$_{3}$
in the fifties,\cite{Jonker56,Gilleo57,Bents57} it has been known
that upon Cr-doping the antiferromagnetic Mott insulator
LaMnO$_{3}$ develops a ferromagnetic component. Moreover, Taguchi
\textit{et al.}~reported (for T $>$ 300 K) that within the
complete concentration range $0 \leq x \leq 1$ the samples show
semiconducting behavior.\cite{Taguchi99} Sun \textit{et al.}~and
Zhang \textit{et al.}~came to similar conclusions for the
temperature range T $<$ 300 K, recently.\cite{Sun01,Zhang00} Due
to the ceramic nature of the samples a direct comparison of the
resistivity values of these studies is vain, but both yield an
increase in resistivity with increasing Cr concentration (at fixed
temperatures) indicating the persistence of an insulating ground
state throughout the whole concentration range. Hence, the
LaMn$_{1-x}$Cr$_{x}$O$_{3}$ system is a promising candidate for
studying the close relationship between JT distortion, SE
interactions and orbital degeneracy, especially because in the
simple ionic picture the Mn$^{3+}$ (3d$^{4}$) ions are partially
substituted by isoelectronic Cr$^{3+}$ ions (3d$^{3}$), which have
the same electronic configuration as Mn$^{4+}$ (e.g.~t$_{2g}^{3}$)
in these compounds. Therefore no mobile charge carriers should be
present in contrast to the Sr-doped case.

However, Sun \textit{et al.}~and Zhang \textit{et al.}~invoked the
possibility of Mn$^{3+}$- O - Cr$^{3+}$ double-exchange
interactions in LaMn$_{1-x}$Cr$_{x}$O$_{3}$ in order to correlate
their electronic transport data with the magnetic properties in
their samples.\cite{Sun01,Zhang00} Magnetoresistance measurements
(T $>$ 77 K) have been reported by Gundakaram \textit{et al.}~in
the system $Ln$Mn$_{1-x}$Cr$_{x}$O$_{3}$ ($Ln$ = La,Pr,Nd,Gd)
indicating the absence of DE interactions.\cite{Gundakaram96}

The influence of Cr-doping in mixed-valence manganites has been
investigated by Cabeza \textit{et al.~}in the system
La$_{0.7}$Ca$_{0.3}$Mn$_{1-x}$Cr$_{x}$O$_{3}$.\cite{Cabeza99a}
Their conclusion, namely that the Cr-ions do not contribute to the
DE mechanism, is in accordance with the observations of Kimura
\textit{et al.~}and Troyanchuk \textit{et al.}, who investigated
the influence of Cr-doping in
Nd$_{0.5}$Ca$_{0.5}$Mn$_{1-x}$Cr$_{x}$O$_{3}$ and
Nd$_{0.6}$Ca$_{0.4}$Mn$_{1-x}$Cr$_{x}$O$_{3}$, respectively.
\cite{Kimura00,Kimura01,Troyanchuk02} In contrast, Sun \textit{et
al.}~again found indications for DE in
La$_{0.67}$Ca$_{0.33}$Mn$_{1-x}$Cr$_{x}$O$_{3}$.\cite{Sun00}

In the present paper we try to shed some light on the
controversial findings by comparing the influence of Cr-doping on
the magnetic properties of LaMnO$_{3}$ with the intensively
studied DE systems La$_{1-x}$(Ca,Sr)$_{x}$MnO$_{3+\delta}$.

\section{Experimental details}

The polycrystalline specimens were prepared using conventional
ceramic techniques. Ultra pure oxide powders (La$_{2}$O$_{3}$
(4N), Mn$_{2}$O$_{3}$ (4N) and Cr$_{2}$O$_{3}$ (4N), Alpha) were
dried, mixed in the appropriate amounts and carefully ball-milled
to ensure homogeneous samples. The LaMn$_{1-x}$Cr$_{x}$O$_{3}$ ($0
\leq x \leq 0.5$) samples have been pressed into pellets and were
prepared by heating in flowing pure nitrogen at 1400 $^\circ$C for
120 hours and then slowly cooled to room temperature. This
procedure was repeated four times. Powder diffraction patterns
were collected employing Cu-K$_{\alpha 1}$ radiation at room
temperature. The magnetic susceptibility and magnetization were
measured using a dc superconducting quantum interference device
(SQUID) magnetometer ($0-50$ kOe, 1.5 K $\leq T \leq$ 400 K).

ESR measurements were performed with a Bruker ELEXSYS E500
CW-spectrometer at X-band frequencies ($\nu \approx$ 9.35 GHz)
equipped with continuous gas-flow cryostats for He (Oxford
Instruments) and N$_2$ (Bruker) in the temperature range between
4.2 K and 680 K. The polycrystalline samples were powdered and
filled into quartz tubes and fixed with either paraffin (at low
temperatures 4 K $\leq T \leq$ 300 K) or NaCl (at 300 K $\leq T
\leq$ 680 K).

\section{Experimental Results}

\subsection{Structural properties}

\begin{figure}[b]
\centering
\includegraphics[width=50mm,clip,angle=-90]{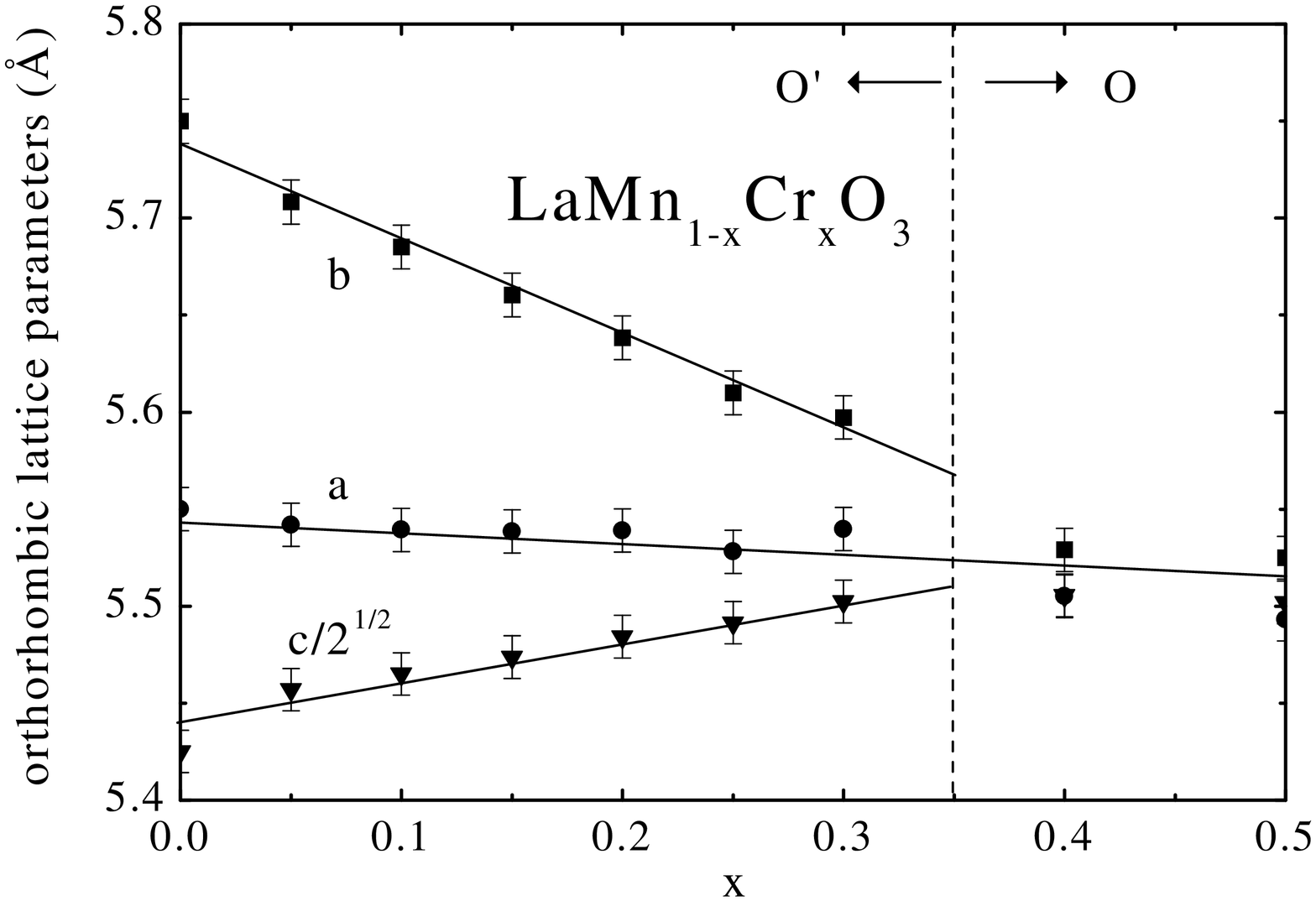}
\vspace{2mm} \caption[]{\label{cellparameters}Orthorhombic lattice
parameters (\textit{a, b, c}) as a function of Cr concentration
$x$. A transition from the Jahn-Teller distorted O$'$-phase to the
pseudocubic ($a \approx b \approx c/\sqrt{2}$) O-phase occurs
between $0.3<x<0.4$. The lines are to guide the eyes}
\end{figure}

In Figure 1 we show the orthorhombic lattice parameters
(\textit{a, b, c}) as a function of the Cr-concentration $x$. All
parameters show a linear dependence on $x$ up to 0.3. A transition
from the Jahn-Teller distorted O$'$-phase to the pseudocubic ($a
\approx b \approx c/\sqrt{2}$) O-phase occurs between $0.3<x<0.4$.
In the Jahn-Teller distorted phase the Mn-O bond lengths in the
MnO$_{6}$ octahedra are highly anisotropic and become isotropic in
the pseudocubic O phase. Our findings are in agreement with the
structural data of Gundakaram \textit{et al.}\cite{Gundakaram96}
The authors also showed that the sample-preparation method
described above ensures the absence of Mn$^{4+}$ ions in these
compounds confirmed by analysis of the samples via iodometric
titration. Therefore we estimate the deviation from the ideal
oxygen value to be less than 1\%. The persistence of the
orthorhombic structure up to $x=0.5$ has also been reported by
Nakazono \textit{et al.}\cite{Nakazono01}

Sun \textit{et al.}~reported that a transition from rhombohedral
to orthorhombic symmetry occurs around $x=0.2$.\cite{Sun01} Since
no detailed analysis of the lattice parameters and information
about the atmosphere used in the preparation procedure was
provided, it is difficult to argue where this discrepancy results
from: A closer look on the X-ray-diffraction spectra presented by
Sun \textit{et al.}~might reveal the persistence of the
orthorhombic distortion or a mixed phase up to $x=0.3$. The
samples in the study by Zhang \textit{et al.}~($x\leq 0.3$)
reportedly were all found to exhibit rhombohedral
symmetry,\cite{Zhang00} whereas Taguchi \textit{et al.}~found
rhombohedral symmetry for $x\geq 0.4$ and orthorhombic for $x\leq
0.4$ and a high Mn$^{4+}$ content due to oxygen non-stoichiometry.
\cite{Taguchi99}

It is known that if upon Sr-doping the percentage of Mn$^{4+}$ in
LaMnO$_{3}$ becomes larger than 20\%, the room-temperature
structure is rhombohedral.\cite{Urushibara95} The influence of
oxygen content on the structural properties has been intensively
studied in LaMnO$_{3+\delta}$ by Prado \textit{et
al.},\cite{Prado99} who found a rhombohedral structure at room
temperature for an oxygen excess of $\delta > 0.09$, which
corresponds to a Mn$^{4+}$ content similar to a Sr concentration
of $x=0.18$. Therefore, it is possible that oxidation
(\textit{i.e.}~oxygen non-stoichiometry) of the samples holds for
the differences in the structural, electronic and magnetic
properties.

Two main reasons account for the different behavior of Cr$^{3+}$
and Mn$^{4+}$ ions (both are not JT active): Firstly, the ionic
radius of the Cr$^{3+}$ ion is larger than the corresponding one
of the Mn$^{4+}$ ion and hence the destabilization of the
Mn$^{3+}$ Jahn-Teller matrix is weaker than in the Sr-doped
samples. Additionally, the mobile nature of the induced charge
carriers (holes) in the case of Sr-doping allows for a dynamical
change of the local electronic configuration and a homogenous
distribution of the Mn$^{4+}$ ions in the JT matrix, whereas the
Cr$^{3+}$ ions are fixed to the lattice sites at the time of
sample preparation.

\subsection{Magnetization}

\begin{figure}[t]
\centering
\includegraphics[width=50mm,clip,angle=-90]{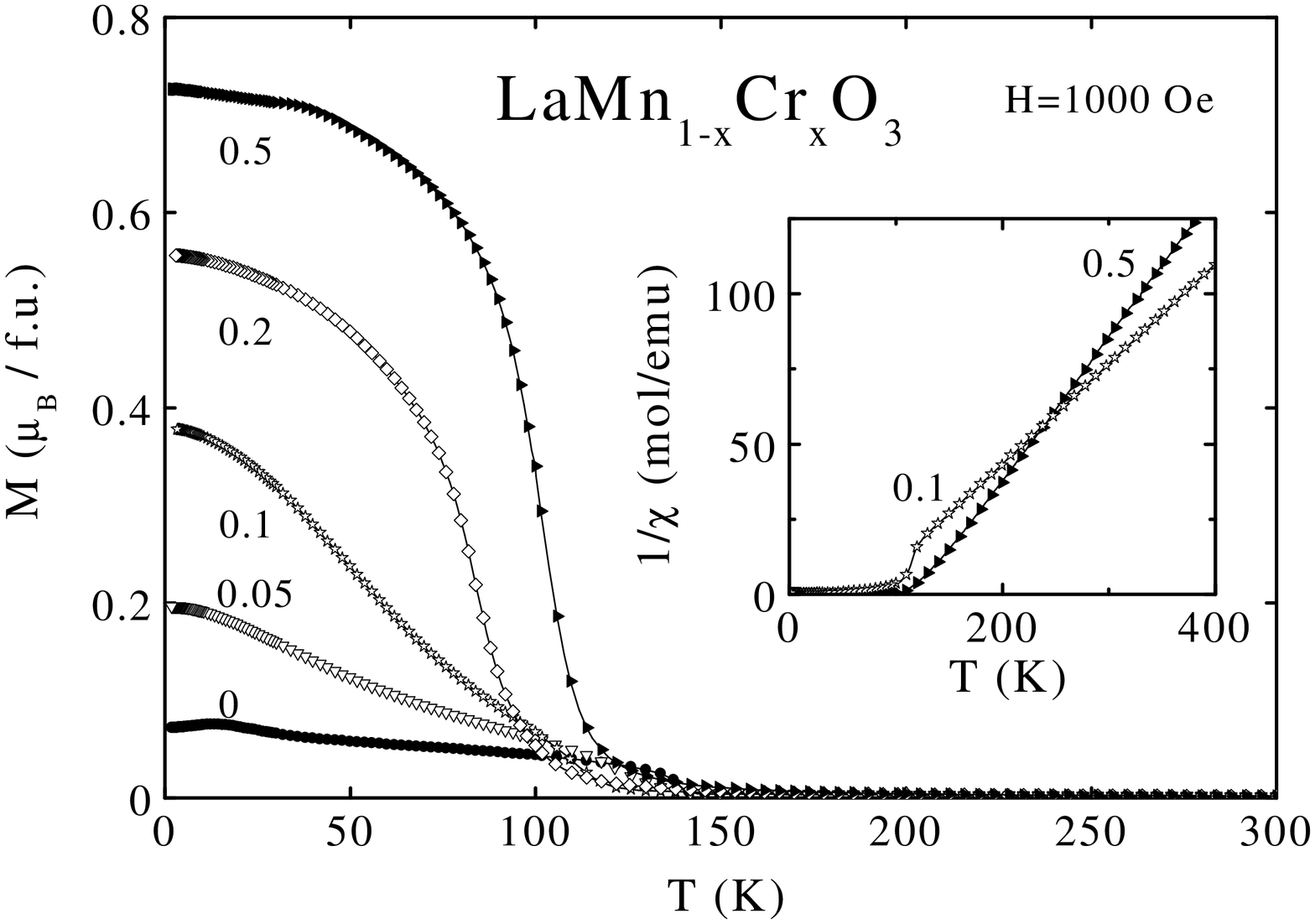}
\vspace{2mm} \caption[]{\label{figure2}Temperature dependence of
the magnetization $M$ with a applied field $H=1000$ Oe. In the
inset the inverse susceptibility $vs.$ temperature is plotted for
$x=0.1$ and $x=0.5$.}
\end{figure}
The magnetization $M$ below the magnetic ordering temperatures
increases upon increasing Cr content, as can be seen from the
temperature dependence of the magnetization in Figure 2. This is
in agreement with the results of Sun \textit{et al.~}and Nakazono
\textit{et al.},\cite{Sun01,Nakazono01} who additionally  observed
a FC/ZFC splitting up to $x=0.5$ indicating a cluster-glass-like
behavior. Since the magnetic transition temperatures cannot be
unambiguously determined from the dc measurements, we restrict our
discussion to the doping dependence of the Curie-Weiss (CW)
temperature $\Theta$. In the inset of Fig.~2 the inverse
susceptibility $vs.$~temperature is plotted for $x=0.1$ and
$x=0.5$.
\begin{figure}[b]
\centering
\includegraphics[width=65mm,clip,angle=0]{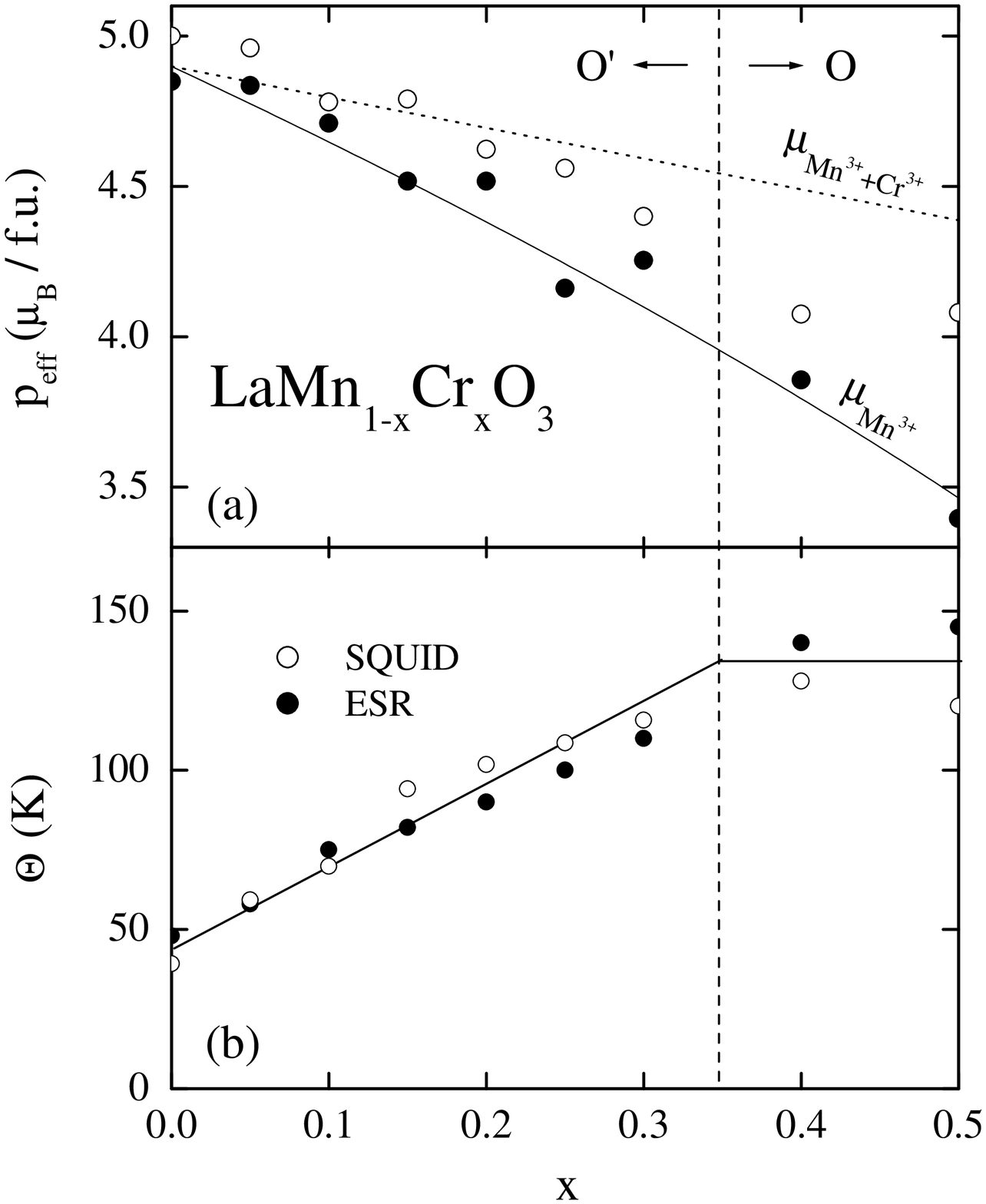}
\vspace{2mm} \caption[]{\label{figure3}Doping dependence of the
effective paramagnetic moments $p_{\mathrm{eff}}$ (a) and the CW
temperature $\Theta$ (b). The lines in (a) show the theoretical
spin-only curve for the effective paramagnetic moment of
Mn$^{3+}$+Cr$^{3+}$ (dotted) and Mn$^{3+}$ (solid line), the lines
in (b) are to guide the eyes.}
\end{figure}
The susceptibilities for all samples follow a CW law
$\sim(T-\Theta)^{-1}$ in the paramagnetic temperature regime. From
this measurements we deduce the effective paramagnetic moments
$p_{\mathrm{eff}}$ (Fig. 3a) and the paramagnetic CW temperature
$\Theta$ (Fig. 3b). From the doping dependence of
$p_{\mathrm{eff}}$ it is noticeable that for $x$ values smaller
than 0.1 the measured values are slightly enhanced with respect to
the simple ionic spin-only values, while for $x \geq 0.2$ they lie
well below the spin-only curve.  Both parameters show a
significantly different doping dependence in the O$'$ and O phases
respectively. While in the Jahn-Teller distorted regime
(O$'$-phase) a linear dependence can be observed up to $x=0.3$,
both $p_{\mathrm{eff}}$ and $\Theta$ become almost constant in the
O phase.

In Figure 4 hysteresis measurements up to 50 kOe at $T=5$ K are
presented. The evolution of a ferromagnetic component upon
increasing Cr-doping is clearly evident. Additionally, upon
increasing Cr concentration the samples become magnetically
softer, as indicated by the decreasing coercive fields in the
hysteresis loops. As shown in the inset of Figure 4, the
spontaneous magnetization $M_{\mathrm{s}}$ (obtained by
extrapolation from the data at the highest magnetic fields, see
lower panel of Fig.~4) increases linearly with the concentration
of Cr$^{3+}$ and reaches a maximum at $x=0.3$. In the O-phase
$M_{\mathrm{s}}$ shows a constant behavior, similar to
$p_{\mathrm{eff}}$ and $\Theta$.

In principle, the behavior of the magnetic parameters discussed
above is in accordance with the early studies by Jonker and
Bents.\cite{Jonker56,Bents57}

\begin{figure}[t]
\centering
\includegraphics[width=65mm,clip]{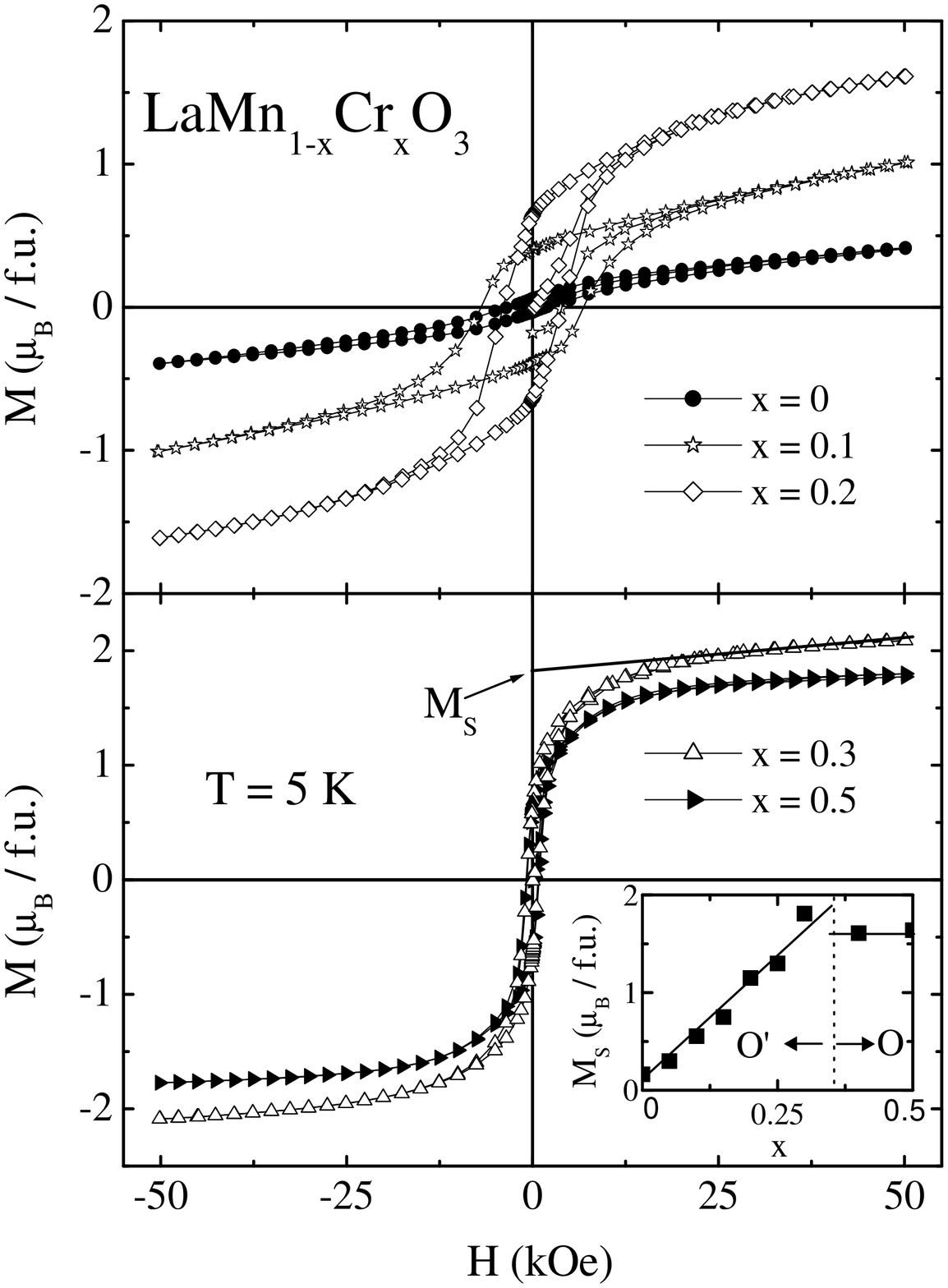}
\vspace{2mm} \caption[]{\label{figure4}Field dependence of the
dc-magnetization $M$ at $T=5$ K. Inset: spontaneous magnetization
$M_{s}$ as a function of Cr-doping.}
\end{figure}

\subsection{Electron Spin Resonance}

\begin{figure}[t]
\centering
\includegraphics[width=60mm,clip,angle=-90]{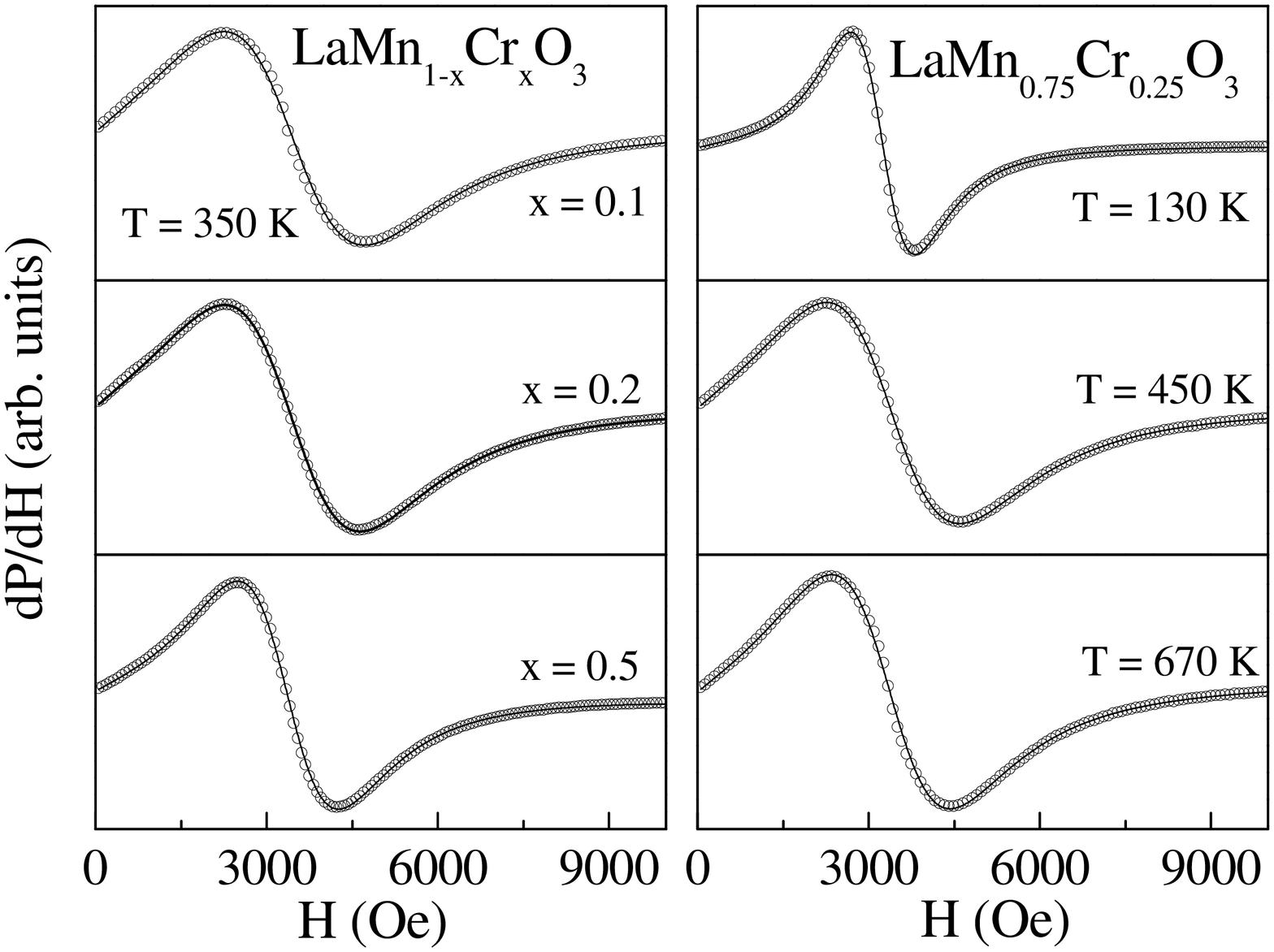}
\vspace{2mm} \caption[]{\label{spectra} ESR spectra of
LaMn$_{1-x}$Cr$_{x}$O$_{3}$. Left column: various Sr
concentrations $x$ at $T = 350$ K. Right column: temperature
evolution of the ESR spectrum for $x = 0.25$. Solid lines
represent the fits using the Lorentzian line shape.}
\end{figure}

Electron spin resonance detects the power $P$ absorbed by the
sample from the transverse magnetic microwave field as a function
of the static magnetic field $H$. The signal-to-noise ratio of the
spectra is improved by recording the derivative $dP/dH$ with
lock-in technique. ESR spectra, which are characteristic for the
paramagnetic regime, are presented in Fig.~\ref{spectra},
illustrating their evolution with Cr concentration $x$ (left
column) and temperature $T$ (right column). Within the whole
paramagnetic regime the spectrum consists of a broad, exchange
narrowed resonance line, which is well fitted by a Lorentzian line
shape as described previously.\cite{Ivanshin00}

The integrated intensity $I(T)$ of the resonance line measures the
spin susceptibility $\chi_{\mathrm{ESR}}$ of the ESR probe. For
ferromagnetically coupled ions its temperature dependence usually
follows a CW law, where $\Theta$ is the CW temperature. Indeed, we
observed a linear behavior of $1/I(T)$ in the paramagnetic regime
within the whole concentration range. The CW temperatures as
determined from the ESR experiments agree well with values
obtained from susceptibility measurements (Fig.~\ref{figure3}(b)).
By comparison with the intensity of the reference compound
Gd$_2$BaCuO$_5$,\cite{Goya96} we obtained the effective moments,
which are in good agreement with the magnetic susceptibility data
except for $x=0.5$.

\begin{figure}[t]
\centering
\includegraphics[width=70mm,clip]{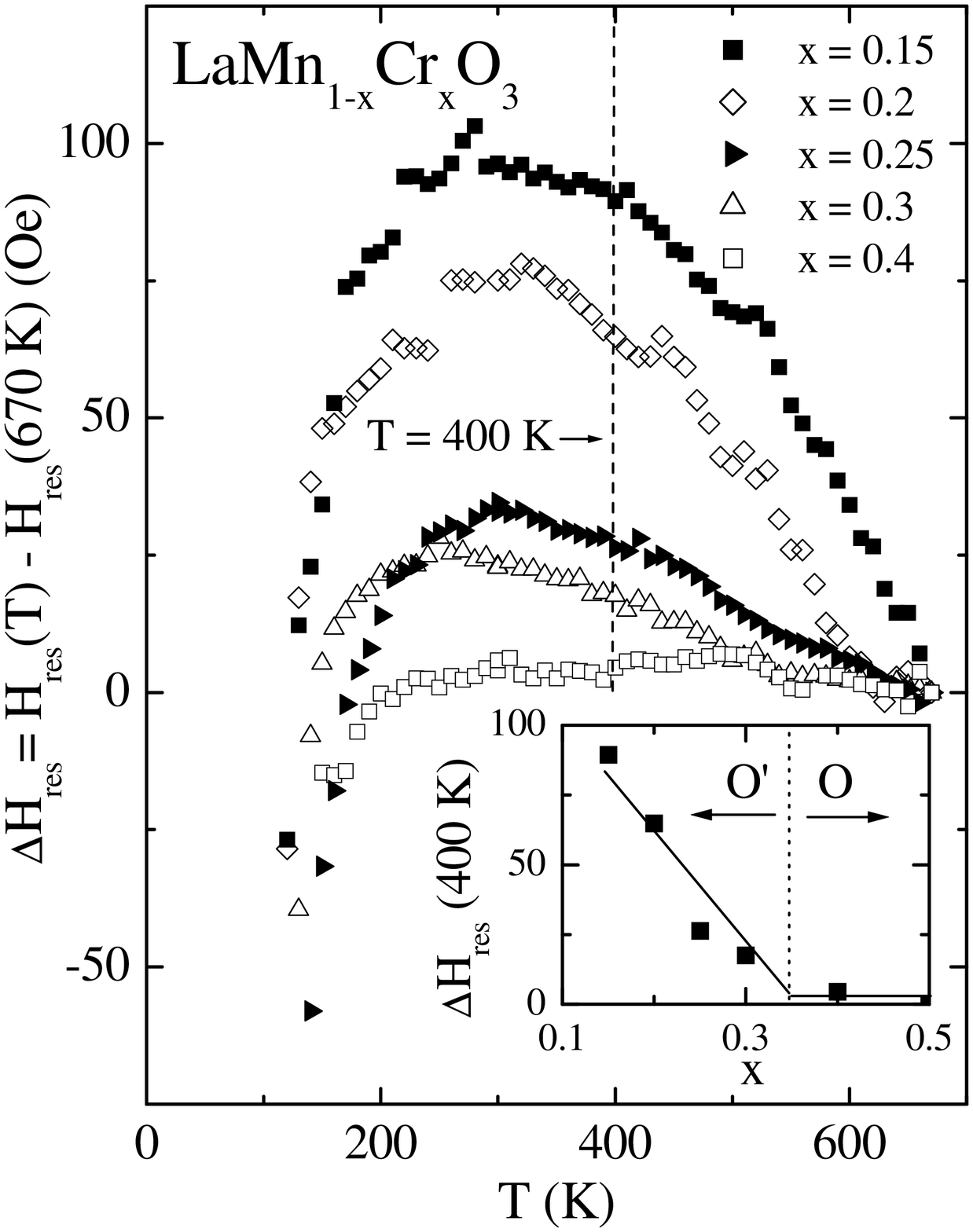}
\vspace{2mm} \caption[]{\label{gvalue}Temperature dependence of
the resonance shift $\Delta H_{res}$ for
LaMn$_{1-x}$Cr$_{x}$O$_{3}$ $(x\geq 0.15)$. Inset: Resonance shift
at 400 K as a function of Cr-doping.}
\end{figure}
In the non-JT distorted orthorhombic O phase $(x=0.4)$ the
resonance field $H_{\mathrm {res}}$ yields an effective g value
$g_{\mathrm {eff}}\approx 1.99$ slightly below the free-electron
value, which is characteristic for transition-metal ions with a
less than half filled d-shell.\cite{Abragam70} In
Fig.~\ref{gvalue} we show $\Delta H_{\mathrm {res}}=H_{\mathrm
{res}}(T)-H_{\mathrm {res}}(670\mathrm{K})$ for concentrations
$x\geq 0.15$. At 670 K the resonance fields for these samples
reach values characteristic for the O phase $H_{\mathrm {res}}(670
\mathrm{K})\approx 3.34$ kOe. For $x< 0.15$ the resonance shift is
even larger, but the structural phase transition seems to take
place beyond our accessible temperature range. However, the
resonance shift at $T=400$ K for $x\geq 0.15$ (see inset of
Fig.~\ref{gvalue}) clearly mirrors the structural phase transition
as the shift decreases with increasing Cr concentration and drops
to almost zero for $x\geq 0.4$.

Finally approaching the ordering temperature from above, the whole
resonance becomes seriously distorted and is strongly shifted to
lower fields due to the internal fields caused by the onset of
magnetic order. A similar evolution of the resonance line has been
reported by Sun \textit{et al.~}and interpreted as an indication
for DE.\cite{Sun01} However, we restrict our discussion to the
paramagnetic phase ($T> T_N$), because in order to characterize
the magnetic order accurately one needs single-crystals of defined
shape instead of randomly oriented powder which gives rise to the
observed distortion of the lineshape.

\begin{figure}[t]
\centering
\includegraphics[width=70mm,clip]{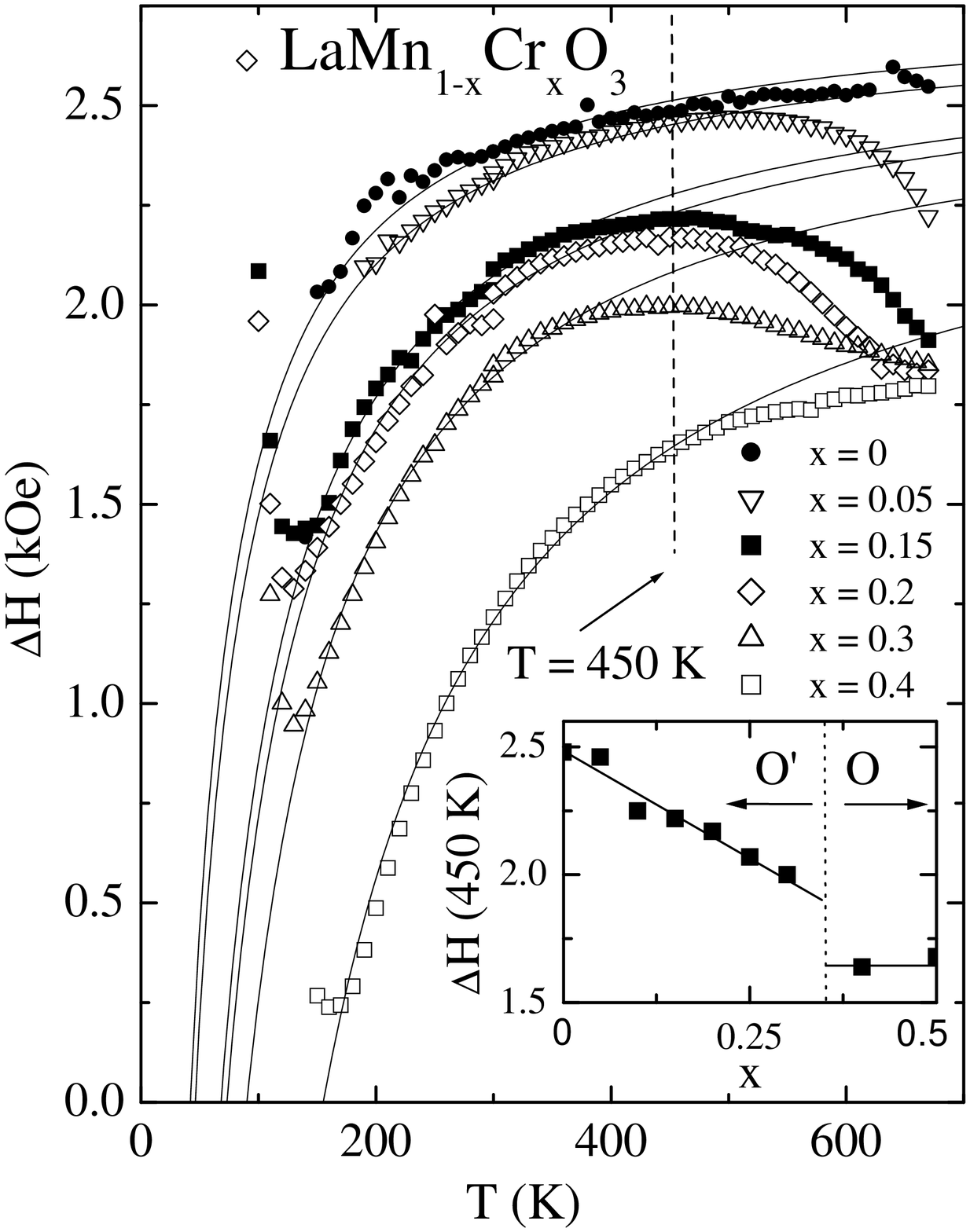}
\vspace{2mm} \caption[]{\label{linewidth}Temperature dependence of
the linewidth for LaMn$_{1-x}$Cr$_{x}$O$_{3}$. The solid lines
represent the fits using eq.~\ref{dhDM}. Inset: Linewidth at 450 K
as a function of Cr-doping.}
\end{figure}
Figure \ref{linewidth} shows the characteristic temperature
dependence of the ESR linewidth. All samples with $x\leq 0.3$ show
a broad maximum in the JT distorted phase and a minimum above the
critical broadening on approaching magnetic order. For the parent
compound LaMnO$_{3}$ the maximum cannot be detected within the
investigated temperature range, but has been reported by Causa
\textit{et al.}~\cite{Causa99} For the highest temperatures under
investigation the linewidth of all samples with $x>0.05$ seem to
converge indicating the range of the high-temperature limit. For
$x>0.3$ the maximum is not observable anymore and the linewidth
shows a monotonous increase with increasing temperatures. The
inset of Fig.~\ref{linewidth} shows the dependence of the
linewidth at 450 K on the Cr concentration. At this temperature
the samples with $x\leq 0.3$ reveal the broad maximum, but have
not yet reached the high-temperature limit. The obvious jump at
the critical concentration reflects the transition from the O$'$-
to the O-phase (Fig.~\ref{cellparameters}) thus revealing a close
correlation between spin relaxation and structural distortion.

\section{Discussion}

\subsection{ESR linewidth and resonance field}

\begin{figure}[b]
\centering
\includegraphics[width=50mm,clip,angle=-90]{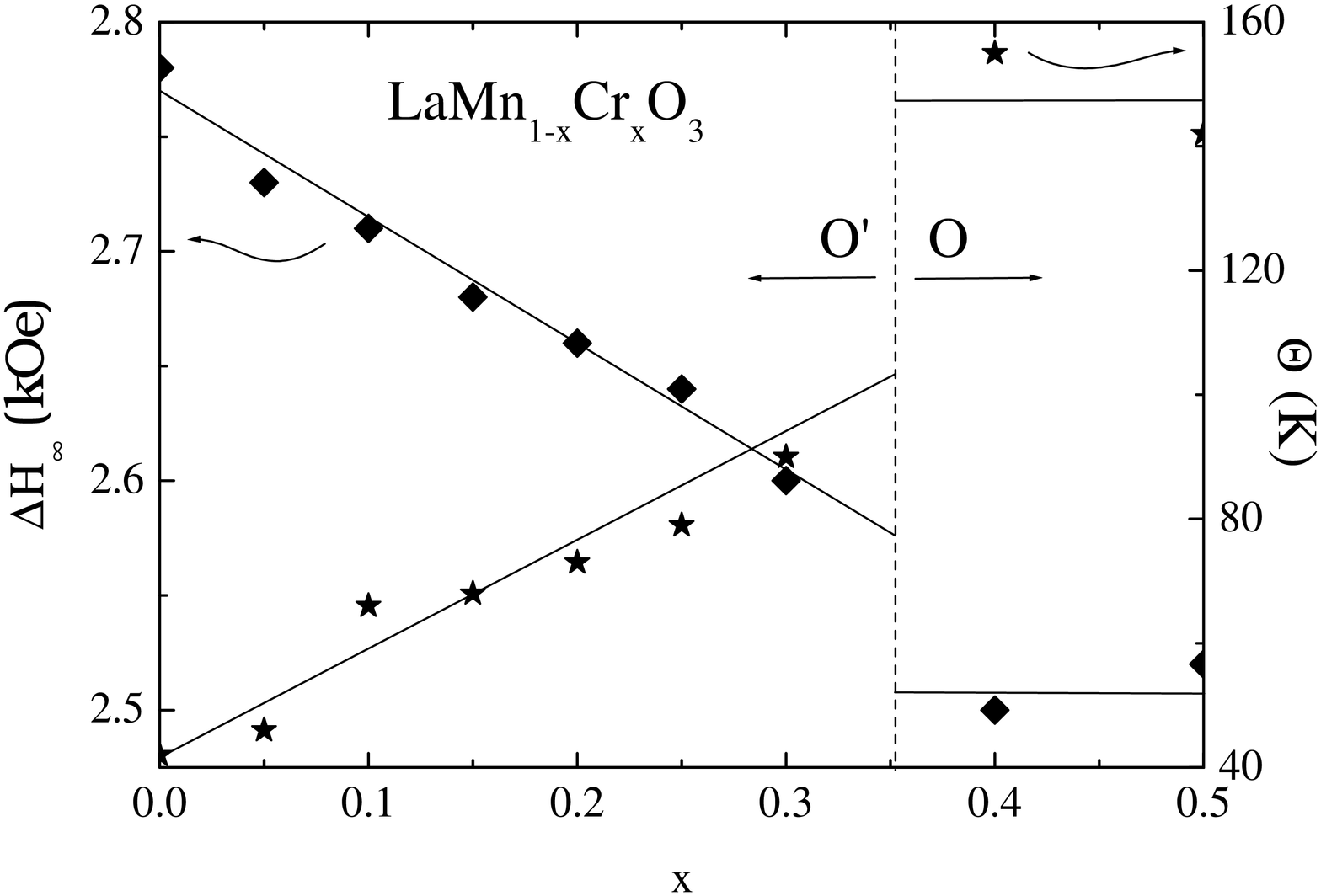}
\vspace{2mm} \caption[]{\label{fitparameters}Doping dependence of
the CW temperature $\Theta$ and $\Delta H_{\infty}$ obtained by
fitting with eq.~\ref{dhDM}, the solid lines  are to guide the
eyes.}
\end{figure}
The temperature dependence of the linewidth and the resonance
field of LaMn$_{1-x}$Cr$_{x}$O$_3$ for $x\leq 0.3$ is very similar
to the ones we observed in La$_{1-x}$Sr$_{x}$MnO$_3$ single
crystals for $x\leq 0.15$.\cite{Ivanshin00} Especially for a Sr
concentration of 5\% the features of linewidth and resonance field
can be compared to the present system.
La$_{0.95}$Sr$_{0.05}$MnO$_3$ is still an antiferromagnetic
insulator ($T_{\mathrm N}=140$ K) dominated by
Mn$^{3+}$-O-Mn$^{3+}$ SE interactions. The presence of the non-JT
active Mn$^{4+}$ ions, however, weakens the cooperative JT matrix
resulting in a reduced JT transition temperature $T_{\mathrm
{JT}}=600$ K. Below this temperature both $\Delta H$ and
$H_{\mathrm{res}}$ show an anisotropic behavior, which can be
regarded as a characteristic feature of the O$'$
phase:\cite{Ivanshin00,Alejandro01}

If the external field is applied within the ferromagnetically
coupled \textit{ab}-plane, the linewidth reveals a broad maximum
within the JT distorted O$'$ phase, whereas $\Delta H$ is nearly
constant, if the external field is parallel to the
antiferromagnetically coupled \textit{c}-axis. We also
investigated a powdered La$_{0.95}$Sr$_{0.05}$MnO$_3$ sample and
found that the linewidth shows the same features as the single
crystal for the external field applied within the $ab$-plane.
Correspondingly, we observed a resonance shift to higher magnetic
fields within the JT distorted O$^\prime$ phase, when the external
field is within the $ab$-plane, and a constant behavior along the
\textit{c}-axis. In our recent approach we presented a detailed
analysis for the angular dependence of the linewidth and the
resonance field in La$_{0.95}$Sr$_{0.05}$MnO$_3$ in the O$'$ phase
by taking into account the crystal field (CF) and the
Dzyaloshinsky-Moriya (DM) interaction,\cite{Deisenhofer02} which
were shown to yield by far the largest contribution to the
linewidth in manganites by Huber \textit{et al.}\cite{Huber99} Due
to the similarities in the temperature dependence of $\Delta H$
and $H_{\mathrm{res}}$ we assume that these two interactions also
account for the ESR properties in Cr-doped LaMnO$_3$.

Unfortunately, in the case of polycrystalline samples such a
detailed analysis is not possible due to the random orientation of
the powder. However, a similar behavior of $\Delta H$ vs.~$T$ has
also been reported in polycrystalline LaMnO$_{3+\delta}$ by Tovar
\textit{et al.},\cite{Tovar99} who found a broad linewidth maximum
in the O$^\prime$ phase up to $\delta =0.04$. The corresponding JT
transition temperatures decrease from 700 K for $\delta =0$ to 400
K for $\delta =0.05$, and for $\delta >0.05$ the cooperative JT
distortion vanishes concomitantly with the broad maximum. The
authors describe the temperature dependence of the
exchange-narrowed linewidth by following Huber \textit{et al.}~as
\begin{equation}
\Delta H(T) = \frac{\chi_0(T)}{\chi(T)}\Delta H_{\infty}
\label{dhDM}
\end{equation}
with the free Curie susceptibility $\chi_0\propto T^{-1}$ and the
static susceptibility
$\chi(T)\propto(T-\Theta)^{-1}$.\cite{Huber99} The
high-temperature linewidth $\Delta H_{\infty}$ depends on the
contributions of both the CF and the DM interaction. Thus, we
fitted the linewidth data for all samples under investigation by
using eq.~1 (see Fig.~7), which can neither describe the critical
broadening of the linewidth in the vicinity of a magnetic
transition nor the structural transition at $T_{\mathrm {JT}}$,
but fits very nicely in between these transitions. The obtained
parameters $\Theta$ and $\Delta H_{\infty}$ are shown in Fig.~8.
Though the obtained CW temperatures are, in comparison with the
values in Fig. 3b, slightly lower for $x\leq 0.3$ (O$'$ phase) and
enhanced for $x> 0.3$ (O phase), the overall tendency is the same.
Moreover $\Delta H_{\infty}$ shows the same behavior as the
linewidth values taken at 450 K (see inset of Fig.~7), confirming
that the maximum linewidth is a good measure for the strength of
the interactions.

The difference in $\Delta H_{\infty}$ observed between the O$'$
and the O-phase is due to the fact that concomitantly with the
disappearance of the cooperative JT effect the contribution of the
single-ion anisotropy of the CF vanishes at $T_{\mathrm {JT}}$
whereas the DM contribution is due to the tilting of the
octahedra, which for LaMnO$_3$ has been shown to remain nearly
unchanged through the transition.\cite{Huang97,Zimmermann01}

The obtained values in both phases are in agreement with those
reported for La$_{1-x}$Ca$_x$MnO$_3$ and LaMnO$_{3+\delta}$ of
$\Delta H_{\infty} \approx 2.7$ kOe for the O$'$-phase and $\Delta
H_{\infty} \approx 2.2$ kOe for the O phase yielding values of the
order of 1 K for both interaction.\cite{Huber99,Tovar99}

The influence of DE on the ESR linewidth in the paramagnetic
regime is not fully understood and still subject of controversial
discussion: Shengelaya \textit{et al.~}found that in
La$_{0.8}$Ca$_{0.2}$MnO$_3$, where the cooperative JT distortion
is already suppressed, the temperature dependence of the linewidth
correlates with the conductivity and both parameters can be
described by a small-polaron hopping model $\sim
1/T\exp(-E_{\mathrm a}/k_{\mathrm B}T)$ with similar activation
energies $E_{\mathrm a}$.\cite{Shengelaya00} In contrast, in the
optimally doped region $x=0.33$ the linewidth has been described
previously by eq.~(\ref{dhDM}) only.\cite{Causa98,Oseroff96} Some
authors argue that due to the different time scales the DE
interaction, which manifests itself in the paramagnetic phase
through the hopping of the JT polaron, cannot have any effect on
the spin relaxation.\cite{Huber99,Rivadulla00} However, we want to
emphasize that in our case the linewidth can be satisfactorily
described without any additional relaxation process connected to
polaron hopping.

\subsection{Magnetic Properties and ESR intensity}
Though all observed parameters show a close correlation to the
structural changes induced upon Cr-doping, the origin of the
observed ferromagnetic component of the magnetization cannot
easily be explained. From the values of the high-field
magnetization (5 T) we deduce that the magnetic phase is not
purely ferromagnetic, since the deviation from the theoretical
values of $3-4 \mu_{\mathrm{B}}/\mathrm{f.u.}$ for the full
magnetic moment of all Mn$^{3+}$ and Cr$^{3+}$ spins cannot be
attributed to temperature induced disorder, only. Therefore, we
have to consider all exchange interactions between the Mn$^{3+}$
and Cr$^{3+}$ ions that could contribute to the magnetic state of
the system:

In the parent compound LaMnO$_3$ the JT distortion lifts the
degeneracy of the $e_g$ orbitals and the Mn$^{3+}$- O - Mn$^{3+}$
SE interaction yields antiferromagnetic coupling along the
\textit{c}-axis and ferromagnetic coupling in the
\textit{ab}-plane. In the case of degenerated $e_g$ orbitals, the
Kanamori-Goodenough rules \cite{Goodenough63} do not apply and the
two-orbital Kugel-Khomskii model yields a purely ferromagnetic
Mn$^{3+}$- O - Mn$^{3+}$ SE interaction.\cite{Kugel82} Due to the
weakening of the JT distortion upon Cr-doping, the degeneracy of
the $e_g$ orbitals can become almost restored (similar to the
ferromagnetic insulating state of La$_{1-x}$Sr$_{x}$MnO$_3$) and
favor ferromagnetism (FM). Taking into account the
Kanamori-Goodenough rules \cite{Goodenough63} the Mn$^{3+}$- O -
Cr$^{3+}$ superexchange is ferromagnetic and therefore can also
account for the increasing values of $\Theta$ in the O$'$-phase
with increasing Cr-doping. Correspondingly, the antiferromagnetic
Cr$^{3+}$- O - Cr$^{3+}$ superexchange should contribute
considerably with increased Cr content and could account for the
stabilization of the magnetic state for $x> 0.3$, where the
observed parameters become almost constant. Finally, a DE
interaction between Mn$^{3+}$- O - Cr$^{3+}$ ions has been
proposed.\cite{Sun01,Zhang00}

Regarding the existence of DE, we can compare our data to
observations of Prado \textit{et al.}~\cite{Prado99} in
LaMnO$_{3+\delta}$, where one can clearly see the impact of
Mn$^{4+}$ - O - Mn$^{3+}$ double exchange on the magnetic
properties: Though the hysteresis loops show a similar behavior
upon increasing oxygen surplus as upon Cr-doping, the magnetic
phase transition from AFM to FM at $\delta\approx 0.04$ is
characterized by the appearance of the full magnetic moment of
$3-4 \mu_{\mathrm{B}}/\mathrm{f.u.}$ These features are also found
upon doping with Sr.\cite{Paraskevopoulos00b} How difficult it is
to distinguish the involved interactions from the dc magnetization
data only can be seen in comparison with the behavior of $M$
vs.~$x$ in electron doped
Ca$_{1-x}$La$_{x}$MnO$_{3}$,\cite{Neumeier00} which looks very
similar to our system. However, this behavior has been found to
correlate with the electron mobility and attributed to the
dynamical charge transfer due to DE interactions. As neither such
correlations nor the appearance of the full magnetic moment has
been observed in LaMn$_{1-x}$Cr$_{x}$O$_{3}$, we will consider the
influence of DE in the following as negligible.

Due to the delicate interplay of the above SE interactions, Jonker
favored the existence of a complex ferrimagnetism
(FiM),\cite{Jonker56} and Bents found in his neutron diffraction
study that not only FM, but also A-type AFM (for $x\leq 0.15$) and
G-type AFM (for $x>0.3$) are present in the system.\cite{Bents57}
A theoretical density-functional study by Yang \textit{et
al.~}based on the data of these early publications supports the
idea of FiM on account of positive Mn$^{3+}$- O - Mn$^{3+}$ and
negative Cr$^{3+}$- O - Cr$^{3+}$ superexchange interactions.
However, positive Mn$^{3+}$- O - Cr$^{3+}$ interaction has not
been found in the calculations.\cite{Yang00} Very recently, Ono
\textit{et al.~}suggested another scenario based on their soft
X-ray magnetic circular dichroism experiments:\cite{Ono00,Ono01}
They found that the Mn ions are aligned parallel to the direction
of the magnetization, whereas the Cr magnetization seems to
disappear. In order to simulate such a behavior by using the Monte
Carlo method, they had to assume a negative Mn$^{3+}$- O -
Cr$^{3+}$ interaction in contrast to the Kanamori-Goodenough
rules. Assuming antiferromagnetic Mn$^{3+}$- O - Cr$^{3+}$
superexchange the latter authors explain the cluster-glass-like
behavior of the system as a frustration of the spins due to the
competing SE interaction.\cite{Nakazono01} Although the reasons
for the disappearance of the Cr magnetization are rather unclear,
such a scenario is in accordance with the observed values of the
spontanous magnetization $M_{\mathrm S}$ values (see inset of
Fig.~4): The linear increase in the JT distorted O$'$ phase can be
attributed to the partially restored degeneracy of the $e_g$
orbitals,\cite{Kugel82} weakening the antiferromagnetic Mn$^{3+}$-
O - Mn$^{3+}$ SE (along the \textit{c}-axis) and finally yielding
only ferromagnetic Mn$^{3+}$- O - Mn$^{3+}$ interaction for
$x>0.3$. Thus, the theoretical values of 1.6 -2
$\mu_{\mathrm{B}}/\mathrm{f.u.}$ are in agreement with our data.

Astonishingly, we also observe that in contrast to Ca/Sr-doped
LaMnO$_{3}$, where the ESR signal has been attributed to both
Mn$^{3+}$ and Mn$^{4+}$ ions, \cite{Causa98,Ivanshin00} the ESR
signal in the paramagnetic regime seems to be originated by the
Mn$^{3+}$ ions only (see Fig.~3): The effective moments determined
by ESR follow the theoretical curve for Mn$^{3+}$ up to $x=0.5$
very nicely. The corresponding values from the magnetization
measurements, however, are slightly enhanced in comparison to the
ESR data, but the overall tendency is similar. In contrast to the
SQUID measurements (see inset of Fig.~2), the ESR experiments were
performed up to 670 K allowing a more reliable determination of
the effective moments. Thus, the frustration of the Cr$^{3+}$ ions
even seems to influence the paramagnetic phase.

\section{Conclusions}

We presented a detailed magnetic analysis of the doping effect of
Cr on LaMnO$_{3}$. The parent compound is an A-type AFM with
orbital order induced by a strong JT effect due to the double
degeneracy of the $e_ {g}$-orbitals of the Mn$^{3+}$ ions. Upon
doping with Cr$^{3+}$ ions, ferromagnetic interactions show up.
The magnetically ordered state can be understood in terms of a
competition of positive Mn$^{3+}$-O-Mn$^{3+}$ and negative
Mn$^{3+}$-O-Cr$^{3+}$ and Cr$^{3+}$-O-Cr$^{3+}$ SE interactions.
Assuming the disappearance of the Cr magnetization the increasing
ferromagnetic component can be ascribed to the Mn$^{3+}$ ions the
partial restoration of the double degeneracy of the $e_
{g}$-orbitals due to the weakening of the JT matrix by non-JT
active Cr$^{3+}$-ions. Additionally, the Jahn-Teller driven
orbital order becomes weaker and finally disappears for doping
levels between 0.3 and 0.4. All parameters, $i.e.$~$\Theta$,
$H_{res}$ and $\Delta H$, reflect this transition in their doping
dependence. Thus, the system can be considered as a ferromagnetic
insulator with respect to Mn$^{3+}$ in the sense of the
two-orbital Kugel-Khomskii model, as the competing negative SE
interactions seem to produce a disappearance of the Cr
magnetization. The ESR properties can be naturally explained by
the structural disorder induced upon Cr-doping and the effects of
the CF and DM interactions. Moreover, the Cr$^{3+}$ ions seem not
to contribute to the ESR signal. From our experimental data we do
not see any need to invoke the possibility of DE interaction
between Cr and Mn ions. The questions about the origin of the
disappearance of the Cr magnetization and why the Cr$^{3+}$ ions
seem to be ESR-silent remain unsolved problems and challenge
further experimental and theoretical investigations.

\begin{acknowledgments}
We want to thank Ch.~Hartinger and A.~Spushk for fruitful
discussions and the authors of Ref.~20 for providing us with a
preprint of their work. This research was supported by the BMBF
via the contract number VDI/EKM 13N6917 and partly by the Deutsche
Forschungsgemeinschaft via SFB 484 (Augsburg).
\end{acknowledgments}


\end{document}